\input epsf
\documentclass[nohyper,notoc]{article} 
\usepackage{epsfig}
\usepackage{bm}
\usepackage{color}
\usepackage[utf8]{inputenc} 

\usepackage{amssymb,amsmath}
\usepackage{mathrsfs}
\usepackage{braket}
\usepackage{ulem}

\def\bit{\begin{itemize}}
\def\eit{\end{itemize}}
\def\ben{\begin{enumerate}}
\def\een{\end{enumerate}}
\def\beq{\begin{equation}}
\def\eeq{\end{equation}}
\def\bea{\begin{eqnarray}}
\def\eea{\end{eqnarray}}
\def\bq{\begin{quote}}
\def\eq{\end{quote}}
\def \lsim{\mathrel{\vcenter
     {\hbox{$<$}\nointerlineskip\hbox{$\sim$}}}}
\def \gsim{\mathrel{\vcenter
     {\hbox{$>$}\nointerlineskip\hbox{$\sim$}}}}
\def\gappeq{\mathrel{\rlap {\raise.5ex\hbox{$>$}}
{\lower.5ex\hbox{$\sim$}}}}
\def\lappeq{\mathrel{\rlap{\raise.5ex\hbox{$<$}}
{\lower.5ex\hbox{$\sim$}}}}

\def\mec{\mu \! \to \! e~ {\rm conversion}}
\def\meg{\mu \to e \gamma}
\def\megg{\mu \to e \gamma \gamma}

\def\meee{\mu \to e \bar{e} e}

\def\LNP{\Lambda_{NP}}

\def\a{\alpha}
\def\b{\beta}
\def\g{\gamma}

\def\m{\mu}

\def\s{\sigma}

\begin{document}

\hspace{131mm} 
OCU-PHYS 519, NITEP 72

\renewcommand{\thefootnote}{\fnsymbol{footnote}}
\begin{center}
{\Large {\bf 
Probing  $\mu  e \g\g$  contact interactions
with $\mu \to e$ conversion}}
\vskip 25pt
{\bf   S. Davidson }$^{1,}$\footnote{E-mail address: s.davidson@lupm.in2p3.fr},
{\bf   Y. Kuno} $^{2,3}$\footnote{E-mail address:kuno@phys.sci.osaka-u.ac.jp },
{\bf   Y. Uesaka} $^{4,}$\footnote{E-mail address:uesaka@ip.kyusan-u.ac.jp },
{\bf  and  M. Yamanaka} $^{5,6}$\footnote{E-mail address:yamanaka@osaka-cu.ac.jp }
\vskip 10pt  

$^1${\it LUPM, CNRS,
Universit\'{e} Montpellier,
Place Eugene Bataillon, F-34095 Montpellier, Cedex 5, France}\\
$^2${\it Department for Physics, Osaka University, Osaka 560-0043, Japan}\\
$^3${\it Research Center of Nuclear Physics, Osaka University, Osaka 567-0047, Japan}\\
$^4${\it Faculty of Science and Engineering, Kyushu Sangyo University, 2-3-1 Matsukadai, Higashi-ku, Fukuoka 813-8503, Japan}\\
$^5${\it Department of Mathematics and Physics, Osaka City University, 
Osaka 558-8585, Japan}\\
$^6${\it Nambu Yoichiro Institute of Theoretical and Experimental Physics 
(NITEP), Osaka City University, Osaka 558-8585, Japan}\\
\vskip 20pt
{\bf Abstract}
\end{center}

\begin{quotation}
  {\noindent\small 
Contact interactions   of  a muon, an electron and two photons  can
contribute  to  the decay $\megg$, but also to   the conversion of 
a  muon  into an electron in the electric field of
a nucleus.  We calculate the $\mu \to e$ conversion
rate,  and show that for  the coefficients of operators involving
the combination $FF \propto |\vec{E}|^2$
(as opposed to  $F\tilde{F} \propto \vec{E} \cdot \vec{B}$), 
the current bound on $\mu \to e$ conversion is more sensitive 
than the bound on  $\mu \to e \gamma \gamma$.

\vskip 10pt
\noindent
}

\end{quotation}

\vskip 20pt 

\setcounter{footnote}{0}
\renewcommand{\thefootnote}{\arabic{footnote}}

\section{Introduction}
\label{intro}

  The observed neutrino masses imply the existence of
contact interactions where charged leptons change flavour.
This is referred to as (Charged) Lepton Flavour Violation (CLFV)
and is reviewed for muon decays in, {\it eg}  \cite{KO}.
Current constraints on several  $\mu \leftrightarrow e$ flavour changing
 processes  are restrictive, and
 experiments under construction \cite{COMET,mu2e,mu3e}
aim to reach  $BR \sim 10^{-16}$.
Some bounds  and future sensitivities  are given in  table \ref{tab:bds}.

 If CLFV is discovered,  experimental bounds on,
or   observations of,  a multitude of  independent processes
would  assist in  discriminating among  models.
This motivates our interest in the less commonly considered
contact interactions involving  a muon,
an electron and two photons. Such interactions
could  mediate  various processes, such
as $\megg$ and  $\mu \to e$ conversion in the electric field
of a nucleus.
The rate for  $\megg$ was calculated by
 Bowman, Cheng, Li and Matis (BCLM)\cite{BCLM},
 whose results are reviewed in section \ref{sec:notn},
 and an experimental  search  with the Crystal
 Box  detector  obtained
 $BR(\megg) \leq 7.2\times 10^{-11}$\cite{CrystalBox}. 
Similar contact interactions,  involving two photons but
Dark Matter instead of leptons,  have been studied
in \cite{Pospelov,Weiner,Uli,OV}.

We will parametrize CLFV interactions   via contact interactions 
involving Standard Model (SM) particles.
This would be appropriate  if the new particles
involved in CLFV are heavy, but  may not
be  generic for  $\mu \to e \g\g$. This decay  could
 be mediated by  $\mu \to e a$ \cite{Jure} followed by $a\to  \g\g$, 
where $a$ is a light (pseudo) scalar such as an Axion-Like Particle \cite{ALPS}.
Recently, the MEG experiment searched for 
collinear photons from this process \cite{MEG2020}.
They found that the branching ratio of 
$\mu^+\to e^+a,a\to \gamma\gamma$ is smaller than 
$\mathcal{O}\left(10^{-11}\right)$ when the mediator $a$ has 
a mass of 20-45MeV and a lifetime below 40 ps.

 In this manuscript, we  calculate the $\mu \to e$ conversion rate
induced by  contact interactions of   $\mu,  e $ and  two photons.
Section \ref{sec:notn}
introduces  the basis of operators (previously given by
BCLM\cite{BCLM}), and   gives their contribution
to $\megg$. The operators are of dimension seven
and eight; we focus on the
dimension seven  operators, which 
can arise from loop corrections to  dimension six
scalar operators. 
Our calculation
of $\mec$ mediated by the $\overline{e} \mu FF$ operator
is presented in Sections \ref{sec:caln} and
\ref{sec:loop},
where we first calculate 
the interaction of the
leptons with the classical electromagnetic field,
then in Section  \ref{sec:loop} find a surprisingly large
``short distance'' loop interaction  of two photons with
individual protons. 
The final discussion section
integrates our  results in the usual expression
for the spin independent Branching Ratio of $\mec$,
and discusses the  current and future  sensitivity to
the  $\overline{e} \mu FF$ operator  coefficients at the experimental scale.  
 Appendix A  considers  loop contributions to the
 $\overline{e} \mu FF$  operator and its relation  to dimension six
 LFV operators in models of heavy New Physics
(and  in passing  mentions an accidental cancellation
 in the contribution of 
 the LFV Higgs coupling  operator ${\cal O}_{EH}$ to $\mec$).
Appendix \ref{app:FA} shows the numerical values of the overlap integral which will be introduced in Section \ref{sec:caln}.

\begin{table}[ht]
\begin{center}
 \begin{tabular}{|l|l|l|}
 \hline
 process & current sensitivity & future  \\
\hline
$\meg $ & $ < 4.2 \times 10^{-13}$(MEG\cite{TheMEG:2016wtm})
 &$  \sim 10^{-14}$ (MEG II\cite{MEGII}) \\
 $\megg $ & $ < 7.2 \times 10^{-11}$(Crystal Box \cite{CrystalBox})
 & \\
$\meee $& $ < 1.0  \times 10^{-12}$(SINDRUM \cite{Bellgardt:1987du}) 
 & $  \sim  10^{-16}$ (Mu3e\cite{mu3e}) \\
$\mu$A$ \to e$A  & 
$< 7 \times 10^{-13}$(SINDRUM II\cite{Bertl:2006up}) &
  $  \sim  10^{-16}$ (COMET\cite{COMET}, Mu2e\cite{mu2e})    \\
    &  &
    $  \sim 10^{-18}$ (PRISM/PRIME\cite{PP})    \\
\hline
\end{tabular}
 \caption{Current bounds on the branching ratios for various 
 CLFV processes, and the expected reach of upcoming experiments.
   \label{tab:bds} }
   \end{center}
 \end{table}


\section{ Notation and Review }
\label{sec:notn}

  A  set of  QED-invariant operators that could  mediate the
decay $\megg$ was given by Bowman, Cheng, Li and Matis (BCLM)\cite{BCLM}:
\bea
\delta{\cal L} &=&\frac{1}{v^3} \left(C_{FF,L}\bar{e}P_L\mu F_{\a\b}F^{\a\b}
+ C_{FF,R}\bar{e}P_R\mu F_{\a\b}F^{\a\b}+
 C_{F\tilde{F},L}\bar{e} P_L\mu F_{\a\b}\widetilde{F}^{\a\b}
+ C_{F\tilde{F},R}\bar{e} P_R\mu F_{\a\b}\widetilde{F}^{\a\b}
\right) \nonumber
\\
&&+\frac{1}{v^4}
 \left(
C_{VFF,L} \bar{e} \g^\s P_L \mu F^{\a\b}\partial_\b F_{\a\s} +
C_{VFF,R} \bar{e} \g^\s P_R \mu F^{\a\b}\partial_\b F_{\a\s} \right.\nonumber\\
&&~~~~~+\left.
C_{VF\tilde{F},L}\bar{e} \g^\s P_L \mu F^{\a\b}\partial_\b \widetilde{F}_{\a\s} +
C_{VF\tilde{F},R} \bar{e} \g^\s P_R \mu F^{\a\b}\partial_\b \widetilde{F}_{\a\s}
	\right) + [h.c.],
\label{L}
\eea
where two changes  have  been made to their notation:  the New Physics scale
in the denominator is taken to be the Higgs vacuum expectation value $v\simeq m_t$, with
$2\sqrt{2} G_F = 1/v^2$ (BCLM took
$m_\mu$), and we use chiral fermions, because this facilitates matching onto
the full SM at the weak scale, and because  the out-going electrons are
relativistic so $\approx$ chiral.

This basis of operators 
is  constructed   to include
all possible Lorentz contractions that give the desired  external
particles (there is  no tensor,
because  there is no  two-index
antisymmetric combination of $FF$ or $F \tilde{F}$ to contract with
$\bar{e}\sigma \mu$\footnote{The tensor operator considered in \cite{Uli}
should vanish.}), so corresponds to a general parametrization of
the interaction at lowest order in a momentum expansion. 
The resulting operators are of
dimension seven and eight.

Curiously, all the operators of eqn (\ref{L})
induce  a  matrix-element-squared for $\mu (P_\m) \to
e(p_e) + \gamma  (k)  + \gamma(q) $
that is proportional to  \cite{BCLM}
$$
|{\cal M}|^2 \propto P_\mu \cdot p_e (k\cdot q)^2 ~~~,
$$
giving a  Branching Ratio 
\bea
BR (\megg) &=& C^2 \frac{2  m_\m^2}{5 v^2} ~~,~~
\label{BR}
\eea
where
\bea
C^2&=& |C_{FF,L} + i \frac{m_\mu C_{VFF,R}}{4 v}|^2  +
|C_{FF,R} + i \frac{m_\mu C_{VFF,L}}{4 v}|^2
+ |C_{F\tilde{F},L} + i \frac{m_\mu C_{VF\tilde{F},R}}{2 v}|^2 \nonumber\\
&&~~~~+
|C_{F\tilde{F},R} + i \frac{m_\mu C_{VF\tilde{F},L}}{2 v}|^2
\eea
The  experimental bound from
Crystal Box \cite{CrystalBox} given in table \ref{tab:bds}
therefore corresponds to
\beq
\frac{m_\m}{v}|C|\lsim 1.3\times 10^{-5} ~~\Rightarrow  |C|\lsim 2.2\times 10^{-2}
~~~.
\label{CBbd}
\eeq

Notice that the BCLM study focuses on ``short-distance''
contributions to $\megg$ mediated by LFV contact interactions.
There could also be a ``long-distance'' contribution, induced  by
a  LFV four-fermion operator that coupled  $\bar{e}\mu$ to the
neutral pion, which decays to $\g\g$. This process could have more
interesting sensitivity to the relevant pseudoscalar  four-fermion
operator  that KTEV \cite{KTEV}(because the muon decays weakly, but the
$\pi_0$ decays electromagnetically), but we do not  consider
it further here because we are interested in $\mu\to e$
conversion, and the $\pi_0$ couples to $\vec{E}\cdot \vec{B}$,
which is negligeably small in the nucleus. 
Consequently, we disregard $\vec{E}\cdot\vec{B}$ contributions in the remainder of this paper.

\begin{figure}[h]
\begin{center}
\epsfig{file=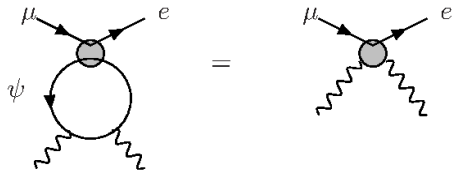,height=3cm,width=8cm}
\end{center}
\caption{Matching  of  scalar  heavy  fermion  operators
 ${\cal O}^{\psi\psi}_{S,XY}$, ${\cal O}^{\psi\psi}_{S,XX}$
onto
the two-photon  operator ${\cal O}_{FF,X}$.
 \label{fig:matchconf}
}
\end{figure}

In many heavy New Physics models, LFV arises at dimension six.
So for a sufficiently high New Physics scale $\Lambda_{NP}$,
it is reasonable to neglect the  dimension $\geq 7$
operators that could be generated at $\Lambda_{NP}$,
because   their contributions to observables
 will be  suppressed by additional factors of
$E_{expt}/ \Lambda_{NP}$. 
However, some of  the operators of eqn (\ref{L}) can
arise at ${\cal O}(1/\Lambda_{NP}^2)$ in a
CLFV New Physics model, with a SM mass scale providing
the additional dimensions in the denominator. 
For  example,  if the  scalar operators
\beq
{\cal O}^{\psi\psi}_{S,XX}\equiv (\bar{e} P_X\mu)(\bar{\psi} P_X\psi)
~~~,~~~
 {\cal O}^{\psi\psi}_{S,XY}\equiv(\bar{e} P_X\mu)(\bar{\psi} P_Y\psi) 
\label{opscaltau}
\eeq
are present in the Lagrangian as
$\delta {\cal L} = \frac{1}{\Lambda_{NP}^2}(
C^{\psi\psi}_{S,XX} {\cal O}^{\psi\psi}_{S,XX}
+ C^{\psi\psi}_{S,XY}{\cal O}^{\psi\psi}_{S,XY})$,
then  at a heavy fermion mass scale  $m_\psi$,  
they  match onto  the two-photon
operator ${\cal O}_{FF,X}$  via the diagram of
figure \ref{fig:matchconf}, with coefficient
\beq
\frac{C_{FF,X}}{v^3} = -\sum_\psi (C^{\psi\psi}_{S,XX}+ C^{\psi\psi}_{S,XY})
\frac{Q^2_\psi N_c \alpha_e}{12\pi m_\psi \Lambda_{NP}^2} 
\label{match1}
\eeq
where $N_c =3$ for heavy
quarks, and is one otherwise. This result
is related to the conformal anomaly\cite{P+S}, and
is the QED version of the matching of  scalar heavy quark  operators
onto  gluons,  performed  by Shifman, Vainshtein and Zakharov \cite{SVZ}
(the $\bar{e}\mu GG$ operators were included in $\mec$ by \cite{CKOT}).

The dimension eight  two-photon operators
${\cal O}_{VFF,X} =
\bar{e} \g^\s P_X \mu F^{\a\b}\partial_\b \widetilde{F}_{\a\s}$
appear more difficult to obtain at ${\cal O} (1/\Lambda_{NP}^2)$.
 Furry's Theorem says that dimension six vector operators, such as
{
$ (\bar{e} \g^\a P_X\mu)(\bar{\psi} \g_\a \psi)$,} do not match onto
 ${\cal O}_{VFF,X}$ via the diagram
 of figure \ref{fig:matchconf}, because an odd number of vector current insertions
 appear on the fermion loop.
 Writing the loop of  figure  \ref{fig:matchconf}
 with external legs amputated and 
  an axial heavy fermion current
{
$ (\bar{e} \g^\a P_X\mu)(\bar{\psi} \g_\a \g_5 \psi)$}
 in the grey blob,
 gives a vacuum  matrix element  that is even under Charge
 conjugation, but odd under CP. Analogously
 to Furry's theorem, it should vanish in
 a CP invariant theory, so we  do not calculate
this diagram in the approximation of CP invariance.

In the next sections, we attempt to calculate the
contribution of  the scalar ${\cal O}_{FF,X}$ operators
to coherent $\mec$. The New Physics scale is
not required to be particularly high : Section \ref{sec:caln}
considers the ``long-range''
classical electromagnetic field of the nucleus,
and should be valid for $\LNP$ 
such that  ${\cal O}_{FF,X}$ is a contact interaction
at the muon mass scale.   Section \ref{sec:loop}
is a QED loop calculation involving protons,
which requires  $\Lambda_{NP}\gg m_p$. 
In Appendix \ref{app:SMEFT}, we will reconsider the
case of $\LNP \gg m_W$.
We do not consider
the contribution of the dimension eight 
${\cal O}_{VFF,X}$ operators;   if  New Physics scale is high,
their contribution to   $\mec$
would be relatively suppressed
 by  ${\cal O}(E^2_{expt}/\Lambda_{NP}^2)$
compared to that of dimension six LFV operators,
and we are not aware of a motivated
light New Physics model that induces these
operators.


\section{The ${\bm \mu \to e}$ conversion rate in the classical electric field}
\label{sec:caln}

We consider the coherent $\mec$ described by the first two terms 
of eqn (\ref{L}). Assuming that the 
muon and the outgoing electron are independently described by 
their  wave functions in a Coulomb potential, the transition matrix is 
\begin{eqnarray}
\mathcal{M}
=
\dfrac{1}{v^3} \displaystyle{\int} d^3r\overline{\psi}_e 
\left( \bm{r} \right) \left( C_{FF,L}P_L+C_{FF,R}P_R \right) 
\psi_\mu^{1s} \left(\bm{r}\right) 
\braket{N|F_{\alpha\beta}F^{\alpha\beta}|N},
\end{eqnarray}
where $\psi_\mu^{1s}$ and $\psi_e$ are respectively  the wave functions of 
a $1s$ bound muon and the outgoing electron.
Here, we omit spin indices for simplicity.
$\ket{N}$ denotes the ground state of a nucleus. 
For an ordinary nucleus, we can safely assume that the electric field 
$\bm{E} \left( \bm{r} \right)$ is spherically symmetric and the 
magnetic field is negligible.

We approximate  the hadronic matrix element 
with a classical field strength as: 
\begin{eqnarray}
\Braket{N|F_{\alpha\beta}F^{\alpha\beta}|N}
=-2 \left\{ E\left(r\right) \right\}^2.
\end{eqnarray}
Diagrammatically, this 
corresponds to assuming  that both exchanged photons
carry three-momentum but no energy (giving a Coulomb
potential), and
neglects excited intermediate states
for the nucleus. It can be compared to the
approximation of  Weiner  and  Yavin   \cite{Weiner}
for Dark Matter scattering on nuclei,
where the nucleus is treated as a  particle
of charge $Z$ in Heavy Quark Effective Theory,
with  a form factor to account for its finite size. \footnote{This 
approach is inconvenient in our case because the
wavefunctions of the electrically charged muon and
electron are easier to include in position space.}

With the amplitude $\mathcal{M}$, the conversion probability 
is given by
\begin{eqnarray}
d\Gamma_{conv.} 
= \frac{d^3p_e}{(2\pi)^32E_e}(2\pi) 
\delta \left(E_e-E_e^{conv}\right) 
\overline{\sum_{spins}}\left|\mathcal{M}\right|^2,
\end{eqnarray}
where the summation includes spin averaging of {{the}}
initial state, and 
$E_e^{conv}$ is the energy of  the signal electron, given by
$E_e^{conv}= \left[ \left( m_N+m_\mu-B_\mu \right)^2 
-m_N^2+m_e^2 \right]/2\left( m_N+m_\mu-B_\mu \right)$. 
Here $B_\mu$ is the binding energy of 
initial muon in the muonic atom. 
The lepton wave functions $\psi_\ell$ ($\ell=e,\mu$) obey the Dirac 
equation in a nuclear Coulomb potential;  our formulation  below
follows \cite{KKO,rose1961}.

For a spherically symmetric potential, 
one can represent the wave function of the bound muon as
\begin{align}
\psi_\mu^{1s}(\bm{r})=
\begin{pmatrix}
G(r)\chi_{-1}^{s_\mu}(\hat{r}) \\
iF(r)\chi_{+1}^{s_\mu}(\hat{r})
\end{pmatrix},
\label{eq:bound_state}
\end{align}
where $\chi$ is a two-component spherical spinor\footnote{
The subscript is the eigenvalue of $\kappa = -\sigma \cdot \vec{L}- 1=\pm
(j+1/2)$, which is $-l-1$ when $\kappa$ is negative, $l$ when
$\kappa$ is positive. Unlike the four-component spinor $\psi$,
the two-component spinors are $L^2$ eigenstates, with different
 $L^2$ eigenvalues in the upper and lower components. See 
\cite{rose1961} for the construction of these states.}. 
The differential equations for the radial wave functions $G(r)$ 
and $F(r)$ are obtained from the Dirac equation as follows, 
\begin{align}
\frac{dG(r)}{dr}-\left(E_\mu+m_\mu+eV_{\rm{C}}(r)\right)F(r)=& 0,
\label{eq:Dirac_eq1_mu} \\
\frac{dF(r)}{dr}+\frac{2}{r}F(r)+\left(E_\mu-m_\mu+eV_{\rm{C}}(r)\right)G(r)=& 0.
\label{eq:Dirac_eq2_mu}
\end{align}
The nuclear Coulomb potential $V_C$ is calculated with 
a nuclear charge density $\rho(r)$ as, 
\begin{eqnarray}
V_C(r) = \int_{0}^{\infty}dr' r'^2 \rho(r') 
\left[ \frac{\theta\left(r-r'\right)}{r}+\frac{\theta\left(r'-r\right)}{r'} \right].
\end{eqnarray}
For the nuclear density, we  adopted two different models,
the two-parameter-Fermi distribution (2pF) and 
three-parameter-Gaussian distribution (3pG), given by
\begin{eqnarray}
\rho_{\text{2pF}}(r) 
=
\frac{\rho_0}{1+\exp\dfrac{r-c}{z}}, ~~~ 
\rho_{\text{3pG}}(r) 
= 
\frac{\rho_0 \left( 1+\omega \dfrac{r^2}{c^2} \right)}
{1+\exp\dfrac{r^2-c^2}{z^2}}. 
\label{Eq:rho}
\end{eqnarray}
The normalization has been used such that 
$Ze=4\pi \displaystyle{\int_0^\infty}\rho(r)r^2 dr$ 
with the normalization factors $\rho_0$ for each type of distribution. 
The parameters, $\omega$, $c$ and $z$, are listed 
in Refs.~\cite{Jager1974,Bellicard1970}.

For simplicity of formulation, we express the wave function of  
the outgoing electron of momentum $\vec{p}_e$ using the partial 
wave expansion: 
\begin{eqnarray}
\psi_e(\bm{r})
=&
\sum_{\kappa,\nu,m} 4\pi i^{l_\kappa} 
(l_\kappa,m,1/2,s_e|j_\kappa,\nu) 
Y_{l_\kappa}^{m*}(\hat{p}_e)e^{-i\delta_\kappa}
\begin{pmatrix}
g^\kappa(r)\chi_\kappa^\nu(\hat{r}) \\
if^\kappa(r)\chi_{-\kappa}^\nu(\hat{r})
\end{pmatrix},
\label{eq:pwe}
\end{eqnarray}
where $j_\kappa$ and $l_\kappa$ are the total and 
orbital angular momentum, respectively. 
We introduced an integer quantum number $\kappa$
 that runs from $-\infty \to \infty$ skipping
 0, and  determines $j$ and $l$ as 
$j_\kappa=\left|\kappa\right|-1/2$ and 
$l_\kappa= j_\kappa+\kappa/2|\kappa|$.
Due to angular momentum conservation, only the waves with 
$\kappa=\mp 1$ contributes to  $\mec$.
$\delta_\kappa$ is a phase shift of the $\kappa$ partial wave, and the incoming boundary condition is taken from 
\cite{rose1961}.
$(l_\kappa,m,1/2,s_e|j_\kappa,\nu)$ is the Clebsch-Gordan 
coefficient, and $Y_{l_\kappa}^{m}(\hat{p}_e)$ is a spherical harmonic.
The radial Dirac equations for each partial wave are
\begin{align}
\frac{dg^\kappa(r)}{dr}+\frac{1+\kappa}{r}g^\kappa(r)-\left(E_e+m_e+eV_{\rm{C}}(r)\right)f^\kappa(r)=& 0,
\label{eq:Dirac_eq1} \\
\frac{df^\kappa(r)}{dr}+\frac{1-\kappa}{r}f^\kappa(r)+\left(E_e-m_e+eV_{\rm{C}}(r)\right)g^\kappa(r)=& 0.
\label{eq:Dirac_eq2}
\end{align}
The normalization of the wave functions is the same as Ref.~\cite{KKO}.

\begin{figure}[t]
	\centering
    \includegraphics[clip, width=8.0cm]{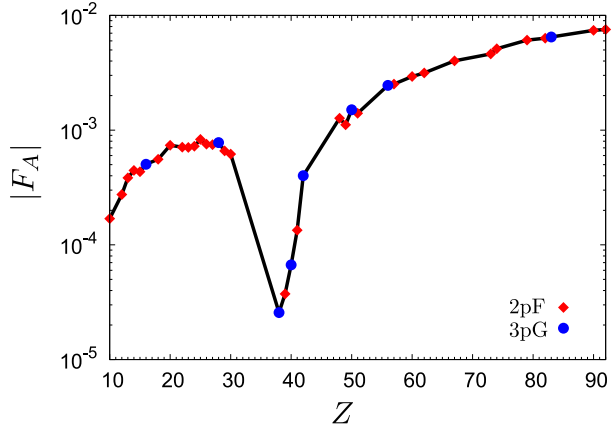}
    \caption{
    $|F_A|$ as a function of atomic number ($Z$) for the target nucleus. 
    Nuclear distributions are 3pG for $Z=16, 28, 38, 40, 42, 50, 56$, 
	and $83$ (blue circle), and 2pF for other nuclei (red diamond).
	}
    \label{fig:FA}
\end{figure}

Then the  conversion probability is 
\begin{eqnarray}
\Gamma_{conv}
= 16G_F^2 m_\mu^5 
\left\{ 
\left|\dfrac{m_\mu}{v}\left(C_{FF,L}+C_{FF,R}\right) F_A^{-} \right|^2
+\left| \dfrac{m_\mu}{v} \left(C_{FF,L}-C_{FF,R}\right) F_A^{+}\right|^2 
\right\}
\end{eqnarray}
where the overlap integrals $F_A^{-}$ and $F_A^{+}$ for 
a target nucleus $A$ are 
\begin{eqnarray}
F_A^{-}
=&
\dfrac{1}{\sqrt{2m_\mu^7}} 
\displaystyle{\int_{0}^{\infty}} dr r^2\left\{E\left(r\right)\right\}^2 
\left\{g^{-1}(r)G(r)-f^{-1}(r)F(r)\right\}, 
\label{FA-}
\\
F_A^{+}
=&
\dfrac{1}{\sqrt{2m_\mu^7}} 
\displaystyle{\int_{0}^{\infty}} dr r^2\left\{E\left(r\right)\right\}^2 
\left\{f^{+1}(r)G(r)+g^{+1}(r)F(r)\right\}.
\label{FA+}
\end{eqnarray} 
Neglecting the electron mass, we have $g^{+1}=-f^{-1}$ and 
$f^{+1}=g^{-1}$, so $F_A^+=F_A^-\equiv F_A$.
For instance, $F_A$ for aluminum ($Z=13$) and gold ($Z=79$) are 
$3.8\times 10^{-4}$ and $-6.1\times 10^{-3}$, respectively. 
$F_A$ for other targets are listed in Appendix~\ref{app:FA}, and 
the absolute values are plotted in Fig.~\ref{fig:FA}.

A few nuclei are modeled by both the 2pF and 3pG distributions, 
in which case we give the results  with the latest distribution. 
Apart from  the dip around $Z=38$ (discussed 
below), different distribution models lead to the same results within 
$\mathcal{O}$(1)\% accuracy. 
The magnitude of $F_A$ continues to grow at large $Z$ (unlike
other overlap integrals \cite{KKO}), because
the squared electric field of heavy nuclei $E(r)^2 \propto 
Z^2$.

\begin{figure}[htb]
	\centering 
    \includegraphics[clip, width=8.0cm]{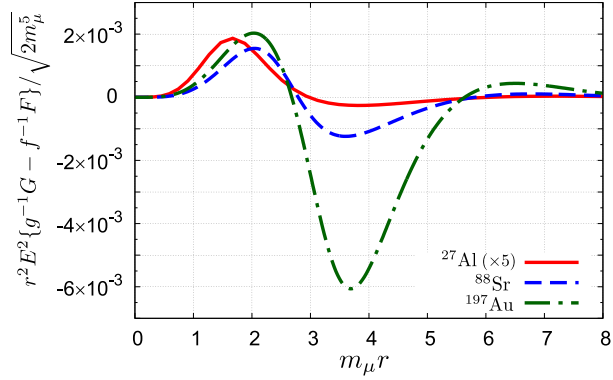}
    \caption{
    Integrand of $F_A$ for $Z=13$, $38$, and $79$. 
	The horizontal axis shows the dimensionless distance 
	from a nuclear center. 
	The amplitude for $Z=13$ is multiplied by a factor $5$.}
    \label{fig:tr_density}
\end{figure}

In Fig.~\ref{fig:FA}, one  sees  a dip in the overlap integral 
in the range $30\lesssim Z\lesssim 50$. In order to interpret
this cancellation,  the integrand of $F_A$  is plotted
as a function of radius in Fig.~\ref{fig:tr_density} for 
$Z=13, 38,$ and $79$. The oscillations arise from 
the electron wave function $g^{-1}$, whose first node is  
at $r\simeq\pi/m_\mu\simeq 5.8$\,fm. 
Since the electric field is maximized around the nuclear surface, 
there is a significant cancellation between the interior and
exterior contributions to the integral when the 
first node of $g^{-1}$  is close to the nuclear radius.  
As a result,  the overlap integral
changes sign at $35 \lesssim Z \lesssim 40$, 
where the nuclear radius is about $5.5$\,fm.

However, the precise prediction of the dip is difficult, 
since the overlap integral at $30\lesssim Z\lesssim 50$ is very 
sensitive to the nuclear model, and some parameters in the 
muonic atom (such as the muon binding energy at 
$\mathcal{O}$(1)\% level). 
In order to reliably predict $F_A$ for these targets, it
would be necessary to model the nuclear 
distributions with  considerable accuracy.

This interesting $Z$ behaviour could be  a signature of the
${\cal O}_{FF,X}$ operators, if their contribution
to the $\mu\to e$ conversion rate is dominant.
For this reason we
exhibit it. However,  as discussed in the next section,
there is a comparable  loop contribution, which arises provided
that the $\mu e \g\g$ interaction remains a contact
interaction  up to scales of a few GeV.

\section{Two-photon exchange with a proton}
\label{sec:loop}

\begin{figure}[ht]
\unitlength.5mm
\begin{center}
\epsfig{file=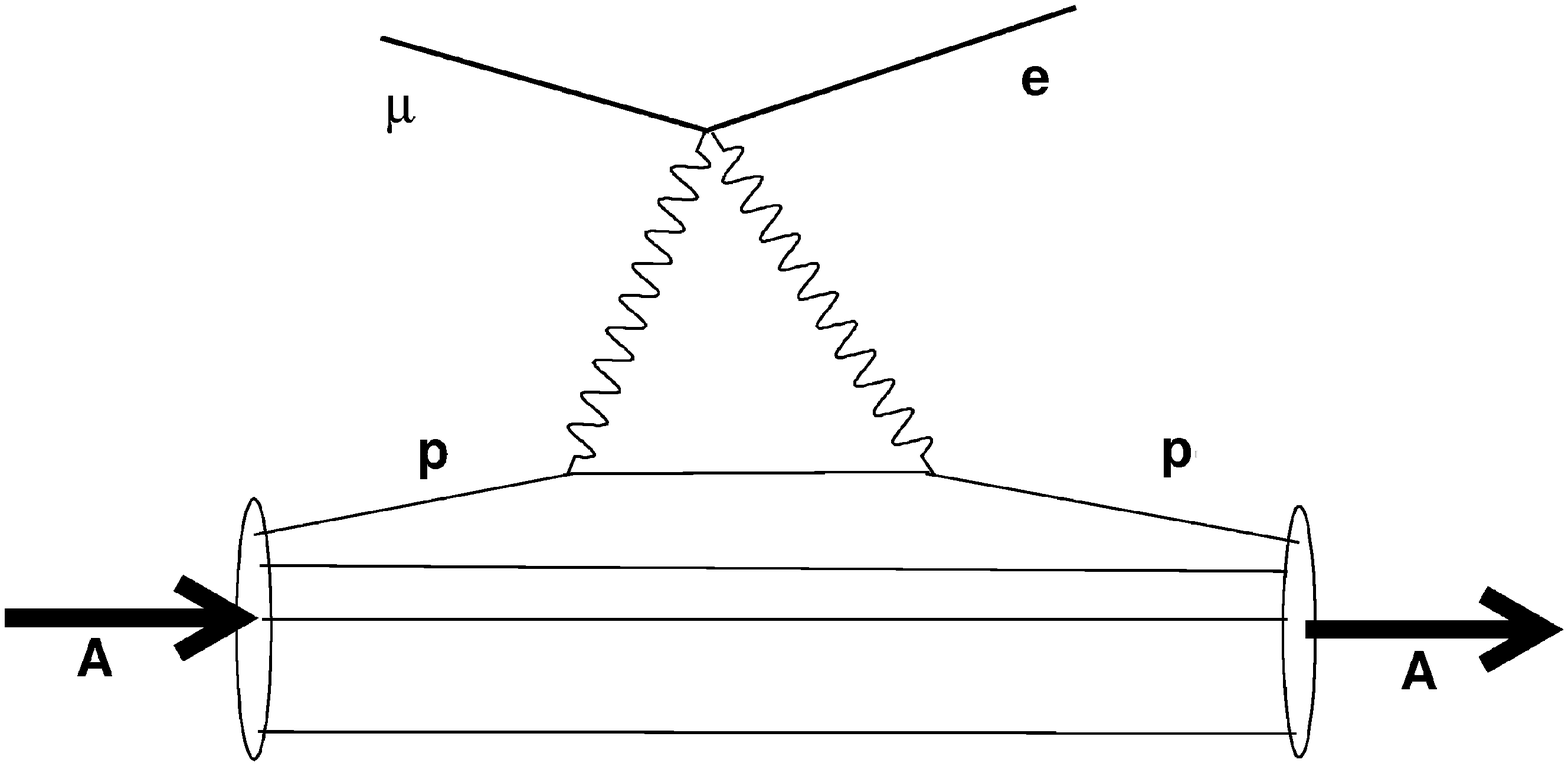,height=4cm,width=8cm}
\epsfig{file=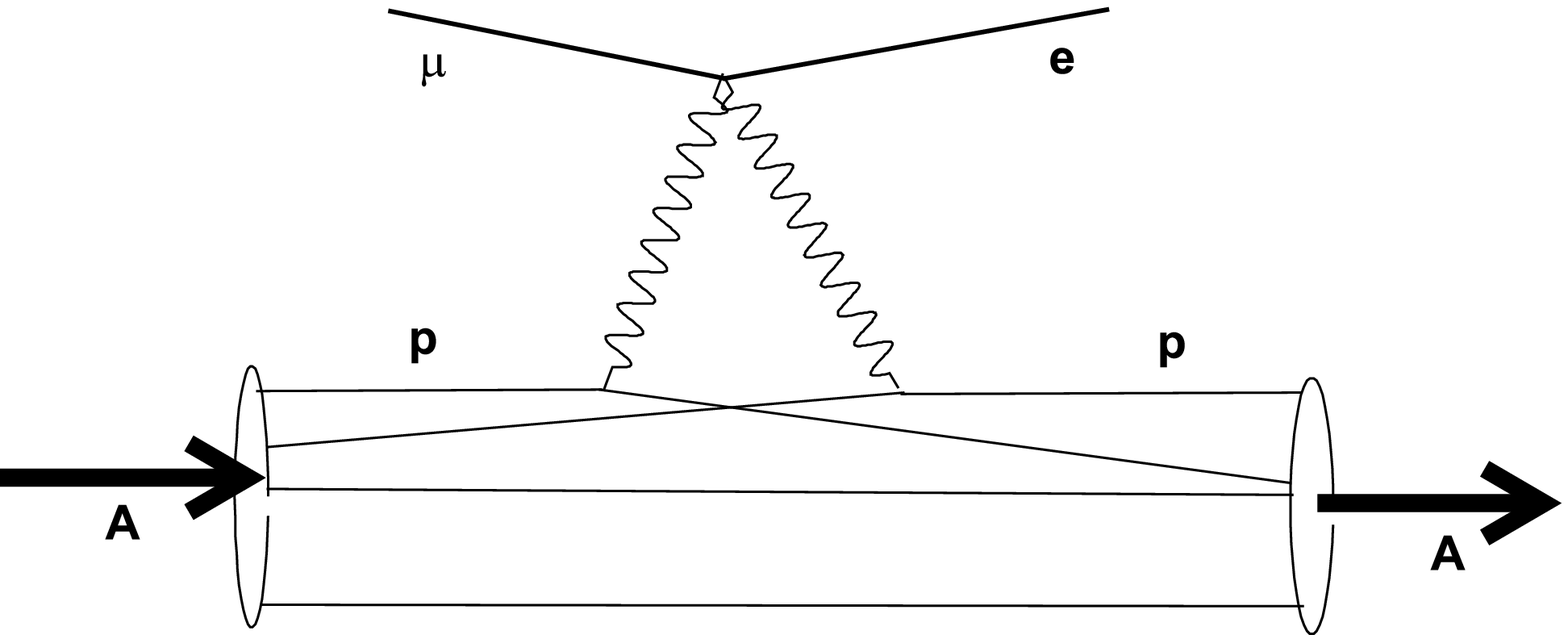,height=4cm,width=8cm}
\end{center}
\caption{  Diagrams for two photons from
the $FF$ operator 
interacting with the nucleus $A$.
\label{fig:ggAA} }
\end{figure}

The nuclear matrix element of the $FF$ operator involves
the expectation value of two nucleon currents, which can
be challenging to calculate. 
If the nucleus is represented as a  non-relativistic
bound state of  protons (and neutrons),  then 
 Wick contractions
give two diagrams for  the interaction of
the ${\cal O}_{FF,X}$ operator with the nucleus,
which are illustrated in Figure \ref{fig:ggAA}.
The sum of both diagrams 
was calculated in the  previous Section \ref{sec:caln},
in the approximation that 
 the protons  remain in their energy levels
 of an external nuclear potential. This implies that
the photons  only carry three-momentum,
so correspond to the Coulomb potential  (which can be
checked in the bound state formalism of
Appendix B of \cite{FHKLX}). This neglects
excited intermediate states of the nucleus,
which are possible  although
the final state nucleus should be in the ground
state, in order to contribute to coherent
$\mu \to e$ conversion.
(We also neglect  correlations
between the two protons, which were considered
in \cite{OV}.)
In this section, we focus on 
the left diagram, where both photons interact with the same proton,
and  estimate the contribution of off-shell photons via
the Renormalisation Group Equations(RGEs) of QED
below the proton mass scale. At
first sight, this diagram appears negligeable, because
it is loop-suppressed ($\propto 1/(16\pi^2)$), and benefits
from  only one factor $Z$ enhancement, as opposed
to $Z^2$ for the tree diagram on the right.

In the RGEs of QED,
the $FF$ operator can mix to   scalar operators
$m_\psi {\cal O}_{S,XX}^{\psi \psi}$,
$m_\psi {\cal O}_{S,XY}^{\psi \psi}$
(defined in  Eqn (\ref{opscaltau})), for $\psi$
a charged point particle.  This
corresponds to the log-enhanced part of the
loop  where both photons interact with the same
proton, can be reliably computed in EFT,
and was considered in \cite{Uli}
for $\psi$ a heavy quark. 
In an EFT of leptons and hadrons below 2 GeV,
we apply this result for $\psi$ a proton,
for scales between 2 GeV and
$m_\mu \simeq$ the momentum exchange
of $\mu\to e$ conversion, which gives
\bea
\Delta C^{pp}_{S,X}(m_\mu) =   -\frac{6\a_{em}m_p}{\pi v}
\ln \frac{2{\rm GeV}}{m_\mu}
C_{FF,X} 
 \simeq - 2.26\times 10^{-4}C_{FF,X}
\label{FFtoSpp}
\eea
where $C^{pp}_{S,X}= \frac{1}{2}(C^{pp}_{S,XL} +C^{pp}_{S,XR})$, and
$C_{FF,X} $ is evaluated at 2 GeV. This
mixing, with $\psi$ a proton,  is discussed
in \cite{OV} but was not included
in \cite{Uli}. {%
It can only be a rough approximation to this
loop, because the not-log-enhanced contributions are
unknown and difficult to estimate. }

We can now calculate the contribution
of  $C_{FF,X} $ to  $\mu\to e$ conversion.
The branching ratio is 
\begin{eqnarray}
BR\left(\mu A \to eA \right)
&=& \frac{32G_F^2 m_\m^5 }{ \Gamma_{cap}}   
 {\Big [ } \big|     
...+C^{pp}_{S,L}  S_A^{(p)}
+ C^{nn}_{S,L} S_A^{(n)} 
+  C_{D,L} {\frac{D_A}{4}} -   \dfrac{m_\mu}{v}C_{FF,L}F_A
 \big|^2   + \{ L \leftrightarrow R \}~ {\Big ]} 
\label{eq:ratio}
\end{eqnarray}
where  $\Gamma_{cap}$ is the muon capture rate 
in a muonic atom \cite{Suzuki:1987jf},  
$S_A^{(N)}$ and $D_A$ are  respectively
the overlap integrals in nucleus $A$ of the
$\bar{N}N$ nucleon current and the dipole operator 
\cite{KKO}, $F_A$ is the overlap integral
for the $FF$ operator from Section \ref{sec:caln},
and ``$...$'' represents the vector coefficients
that we do not discuss. If the principle source
of $\mu\to e$ flavour change at a
scale of  2 GeV is  ${\cal O}_{FF,X}$, then
on  aluminum and gold, we have
\begin{eqnarray}
\frac{BR\left(\mu A \to e A \right)}
{\left|C_{FF,L}\right|^2+\left|C_{FF,R}\right|^2}=
\begin{cases}
6.6 \times 10^{-9} |1+15|^2 & \text{ for }^{27}\mathrm{Al}  \\
9.1 \times 10^{-8} |-1 + 3.8|^2 & \text{ for }^{197}\mathrm{Au} 
\end{cases}.
\end{eqnarray}
where between the absolute values is first the tree
contribution, then the loop. Unexpectedly, the loop contribution 
could be larger than the tree for light and heavy nuclei. 

Let us briefly discuss how  this can occur.
Naively, the loop amplitude should be  suppressed
relative to the  tree contribution 
by $1/(16 \pi^2 Z)$. However:
\bit
\item the  numerical factor from the loop is large:
Eqn (\ref{FFtoSpp}) is $\sim  2\a \log$, 
rather than being $\sim \frac{\a}{4\pi} \log$.
\item   the classical amplitude is suppressed by $1/(4\pi)$,
because  the electric field of a point charge $Z$ is 
$|\vec{E}(r)| =Ze/(4\pi r^2)$,
so a factor $4\pi$ remains in the denominator
when $E^2$ is integrated over the volume of the nucleus.
Combined with the  first effect,  this compensates the $1/16\pi^2$ suppression.

\item the $FF$ operator is of dimension seven, so the
amplitude is proportional to an energy scale. For
the   loop, this is the proton mass,
whereas for the classical process, it is a combination of
the momentum transfer ($m_\mu$) and the inverse 
nuclear radius, which  turns out to be $\sim m_\mu/\pi^2$.
So this ratio of energy scales (over)compensates the
$Z$ suppression of the loop.
\eit

Alternatively, the second point (and part of the third),
can be seen by noticing 
that the  overlap integral $S_A^{(p)}$  is  large   compared 
to  $F_A$.
For simplicity, we assume a uniform proton distribution 
$\rho \propto Z \Bigl( \dfrac{4\pi}{3} R^3 \Bigr)^{-1}$ for 
a nuclear radius $R \sim 1.1 A^{1/3} \,\rm{fm}$.
Since the nuclear electric field is maximized around the nuclear 
surface, we approximate the electric field as one at the surface, 
$|\vec{E} (r)| \simeq Ze/(4 \pi R^2)$.
Hence, the ratio of the overlap integrals is $\left|F_A/S_A^{(p)}\right| \simeq 
2m_\mu^{-1} \Bigl[ \dfrac{Ze}{4\pi R^2}  \Bigr]^2/ 
\Bigl[ Z \Bigl( \dfrac{4\pi}{3}R^3 \Bigr)^{-1} \Bigr] =2Z\alpha/(3m_\mu R)
\sim 0.02$ for $^{27}$Al ($0.06$ for $^{197}$Au), where the overall factor 
$2m_\mu^{-1}$ covers the typical scale of $\mu \to e$ conversion and 
the difference of normalization for overlap integrals between $F_A$ and 
$S_A^{(p)}$ \cite{KKO}. 
That is naive understanding that the overlap $F_A$ is small compared 
to $S_A^{(p)}$. The numerical calculation tells us that the ratio is 
$0.02$ for $^{27}$Al ($0.1$ for $^{197}$Au).

\begin{figure}[htb]
	\centering
    \includegraphics[clip, width=8.0cm]{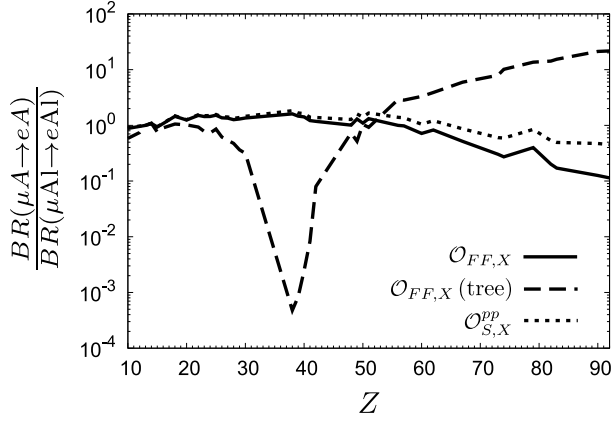}
    \caption{
    Branching ratios for $\mu \to e$ conversion	normalized to that for aluminum.
    The solid line shows the result for $\mathcal{O}_{FF,X} = \bar{e} P_X \mu 
	F^{\alpha \beta} F_{\alpha \beta}$, where we consider two contributions:
	one is the interaction with the classical electric field of the nucleus, and the other is the effect of loop mixing of $\mathcal{O}_{FF,X}$ into $\mathcal{O}_{S,X} = (\overline{e} P_X \mu)(\overline{p} p)$	due to two-photon exchange.
	If neglecting the loop contribution, we obtain the dashed line.
	For comparison, the dotted line gives the normalized branching ratio for only the scalar operator $\mathcal{O}_{S,X}$.
    \label{fig:BR}}
\end{figure}

Figure \ref{fig:BR} shows the branching ratios for targets of 
atomic number $Z$ normalized by that for aluminium. 
The branching ratio via the scalar CLFV operator, 
$\mathcal{O}^{pp}_{S,X} = \bar{e}  P_{X} 
\mu (\bar{p} p)$, is also shown to highlight the difference 
of $Z$ dependence. 
Two features of the overlap integral $F_A$
can be seen: first, it has an additional factor of  $Z$,
due to the extra $F^{\mu \nu}$, and second, it
becomes negative at large $Z$.
The first point is illustrated
by the dashed line, showing
the branching ratio  induced only by
the tree contribution of the  $\bar{e}\mu FF$ operator,
which continues to increase at large $Z$. This  differs from 
the high-$Z$ falloff of the branching ratios
due to  the familiar dipole, scalar or vector 
operators \cite{KKO}.
The solid line includes the
tree and loop contributions of 
 the  $\bar{e}\mu FF$ operator,
which interfere destructively at large
$Z$, where $F_A$ is negative but  the scalar overlap
integral is positive. This sign difference,
combined with the  increasing magnitude of
$F_A$  at large $Z$, causes the branching ratio
to decrease  for increasing  $Z\gsim 50$.
The shape and magnitude of this feature
differ  from  the high-$Z$ decrease 
 of dipole, scalar or vector 
operators \cite{KKO}.
We stress that this feature could be used to discriminate 
 the  $\bar{e}\mu FF$ operator from other CLFV operators.

\section{Summary}
\label{sec:disc}

In this manuscript, we calculated the contribution of low-energy
$\bar{e}\mu\g\g$ contact interactions to $\mu \to e$ conversion on nuclei.
We  considered the
first two operators of eqn (\ref{L}), which are CP-even, of dimension seven, and
involve $F_{\mu\nu} F^{\mu\nu} \supset |\vec{E}|^2$. Other possibilities are
discussed in section \ref{sec:notn}.

If the $\mu e \g\g$ interaction
is a contact interaction at momentum transfers $\sim m_\mu$,
then  there is a contribution to $\mu \to e$ conversion
from the leptons interacting with the electromagnetic field of the nucleus. 
The calculation  is outlined in section \ref{sec:caln}.
It  relies on the overlap integrals in the nucleus,
of the electron and muon wave functions with
the electric-field-squared, which are given in eqns (\ref{FA-},\ref{FA+}).
{%
This contribution has an interesting and rare
feature : it changes sign
at intermediate $Z$ (= the electric charge of
the target nucleus). }

If the $\mu e \g\g$ interaction
remains  a contact interaction at larger momentum transfers $\gsim m_p$,
then the dominant contribution of ${\cal O}_{FF,X}$  to $\mec$  
 arises  from loop mixing  into the
scalar proton operator ${\cal O}_{S,X} = (\bar{e}P_X \mu)(\bar{p} p)$,
as discussed in Section \ref{sec:loop}. Naively, the loop
amplitude  is suppressed by $1/(16\pi^2 Z)$,
but overlap integrals, energy ratios, and numerical factors 
more than compensate, as discussed at the end of the section.  
{%
The  combined tree and loop contributions exhibit  a unique
$Z$-dependence  that could be used to distinguish the
$\overline{e}P_X \mu FF$ operator from other
operators. }
The Branching Ratio for $\mec$ induced by ${\cal O}_{FF,X}$ is
given in Equation (\ref{eq:ratio}), and plotted in Figure \ref{fig:BR}.

If the Branching Ratio for  Spin Independent
$\mec$\cite{KKO} is expressed  as a function of
operator coefficients at a scale of 2 GeV,  our
results for ${\cal O}_{FF,X}$  can be included as
\bea
BR(\mu A \to e A)\!\! &=& \!\!\frac{32 G_F^2m_\mu^5}{\Gamma_{cap}}\!\!\!\sum_{X\in\{L,R\}}\!\!{\Big |}
C_{D,X}\frac{D_A}{4} +(9.0C^{uu}_{S,X}  + 8.2C^{dd}_{S,X}  + 0.42C^{ss}_{S,X})S^{(p)}_A
 \nonumber\\
&&
+(8.1C^{uu}_{S,X}  + 9.0C^{dd}_{S,X}  + 0.42C^{ss}_{S,X})S^{(n)}_A
+ ... \nonumber\\
&&- C_{GG,X} \frac{8\pi m_N}{9\a_s(2m_N) v}
(0.90 S^{(p)}_A + 0.89 S^{(n)}_A)  - C_{FF,X} \left(\frac{ m_\m}{ v} F_A
+ \frac{18 \a m_p}{\pi v} S_A^{(p)}\right)
{\Big |}^2~~~~~
\label{BR1}
\eea
{%
where
 $D_A$   and $S^{(N)}_A$ are the overlap integrals inside the
nucleus  $A$,   with respectively the
electric field  or the appropriate
nucleon ($N \in \{n,p\}$) distribution,
which can be found in \cite{KKO}.
$\Gamma_{cap}$ is the muon capture rate
on nucleus $A$ \cite{Suzuki:1987jf},
$C_{D,X}$ is the dipole  coefficient,
  $\{C^{qq}_{S,X}\}$ are the coefficients
of $2 \sqrt{2} G_F(\bar{e} P_X\mu)(\bar{q} q)$,
and
 the ``$+...$''
represents  the contributions
of vector operators involving a light quark bilinear.
This expression uses
 the quark  densities in the nucleon of
References\footnote{
These are the ``EFT'' determinations, which are
$\sim 50\%$ larger than the lattice results \cite{LLF}.}
\cite{Hoferichter:2015dsa,Alarcon:2011zs,Alarcon:2012nr,Junnarkar:2013ac},   
 the gluon density   \cite{SVZ,CKOT}
\beq
\langle N | GG(x)|N \rangle \simeq  \frac{8\pi m_N}{9\a_s(2{\rm GeV})}
\langle N | \overline{N} N(x)|N \rangle~~,
\eeq
and the last term gives  the
contribution  of the operators
${\cal O}_{FF,X}$, at tree level via the overlap integral $F_A$ 
tabulated  in  Appendix \ref{app:FA}, and via  one
 loop mixing to the scalar proton density
 ($\ln(2\,\rm{GeV}/m_\mu) \simeq 3$ is used in the last term).

}

The SINDRUM II experiment searched for
$\mec$ on gold, and obtained the
upper bound   $BR(\mu {\rm Au} \to e  {\rm Au}) \leq 7\times 10^{-13}$
\cite{Bertl:2006up}.
If we assume that only the ``gauge boson operator''  coefficients
are non-zero (at a scale of  2 GeV), 
this corresponds to the bound:
\bea
 4.9 \times 10^{-8}&\gsim&  {\Big|}
0.222 C_{D,X}  
 -0.038 C_{GG,X}  - 4.8 \times 10^{-5}  C_{FF,X}
{\Big|}~~~,
\label{bdnum}
\eea
which gives,  in the absence of
$C_{D,X}$ and $ C_{GG,X}$,
\beq
 |C_{FF,X}|\leq 1.0 \times 10^{-3} ~~~.
\label{SINDRUMbd}
\eeq
This is a better sensitivity than  that given in
eqn (\ref{CBbd}) from the Crystal Box search for $\megg$.
Searching for $\megg$ nonetheless remains an
interesting and complementary channel, because
it probes all the operators of eqn(\ref{L}).
 Experimental constraints 
on $\mu\bar{e}\g\g$  coefficients are summarised
in table \ref{tab:bd2s}.
\begin{table}[ht]
\begin{center}
 \begin{tabular}{|c|l|c|}
 \hline
coefficient & constraint & process  \\
\hline
$|C_{FF,X} +  {im_\mu C_{VFF,Y}}/{(4 v)}| $ & $ < 2.2 \times 10^{-2}$
 &$ BR(\megg)< 7.2\times 10^{-11} $ \\
 $|C_{F\tilde{F},X} +  {im_\mu C_{VF\tilde{F},Y}}/{(4 v)}| $ & $ < 2.2 \times 10^{-2}$
 &$  BR(\megg)< 7.2\times 10^{-11}$ \\
$|...+C_{FF,X}|$ & $ <  1.0 \times 10^{-3}$
 & $BR(\mu \, {\rm Au}\! \to e\, {\rm Au})< 7\times 10^{-13} $ \\
 \hline
\end{tabular}
 \caption{ $\mu\bar{e}\g\g$ operator coefficients 
 bounded by  $\megg$ 
 \cite{CrystalBox}, and the
 sensitivity of $\mu \textrm{Au} \to e \textrm{Au}$\cite{Bertl:2006up} obtained in this manuscript.
 The operators are given in eqn (\ref{L}), $\{X,Y\} \in \{L,R\}$
 with $X\neq Y$ and $v=174$ GeV.
   \label{tab:bd2s} }
   \end{center}
 \end{table}

The upcoming COMET and Mu2e experiments plan to  start with an
Aluminium target. Combining eqn (\ref{BR1})  and 
$F_{Al} = 3.8 \times 10^{-4}$, we obtain a future sensitivity  of
\beq
 |C_{FF,X}|\leq 7.6 \times 10^{-6} 
\left(\frac{BR(\mu {\rm Al} \to e  {\rm Al})}{10^{-16}}\right)^{1/2},
\label{future}
\eeq
The sensitivity to $C_{FF,X}$  would be improved
by  two orders of magnitude with  an expected branching ratio
of $\sim 10^{-16}$ on the light target Aluminium.

Finally, we comment on the interest of  the $(\bar{e} P_{L,R} \mu) FF$
operators in  identifying heavy New Physics in the lepton sector. 
These operators are of dimension seven in the QED$\times$QCD-invariant
EFT below $m_W$, and dimension eight above. However,  they
can be mediated by not-so-heavy, feebly coupled pseudoscalars of
mass $m\gg m_\mu, m_p$, and  in the case of New
Physics at scales $\gg m_W$, they can 
be induced  in matching out heavy fermion scalar operators of dimension six, as
illustrated in figure \ref{fig:matchconf}, and given in eqn (\ref{opscaltau}).
However, the dominant contribution of such  scalar  operators
to $\mu \to e$ conversion arises via  the
dipole or $(\bar{e} P_{L,R} \mu)GG$ operators.
In  Appendix A, we  estimate  the sensitivities
of  $\meg$ and $\mu \to e$
conversion to  scalar  operators.


\subsection*{Acknowledgements}

We thank Jure Zupan and Lorenzo Calibbi for relevant comments
on a first version of this manuscript. 
SD is happy to  thank Vincenzo Cirigliano,
Martin Gorbahn,  Martin Hoferichter, Marc Knecht  and Aneesh Manohar
for useful conversations. We thank the FJPPL for funding to  HEP-06, which made this manuscript possible.
This work was supported by JSPS KAKENHI Grant Numbers JP18H05231 (YK) and JP18H01210 (YU).

\appendix


\section{$\bar{e} \mu FF$  in the RGEs}
\label{app:SMEFT}

In this appendix, 
we briefly consider how a heavy
New Physics model could
induce the dimension seven $FF$ operators.
Parenthetically, we notice a partial
cancellation in the contribution of
LFV Higgs interactions to $\mec$.

\subsection{Obtaining $\bar{e} \mu FF$ at ${\cal O}(1/\LNP^2)$}

We  assume that New Physics    generates
dimension six CLFV operators  at  some  scale $\LNP >m_W$,
which is high enough that dimension $\geq 7$
operators can be neglected.  Then
${\cal O}_{FF,X}$ can be generated in matching out
dimension six operators, such as 
a Higgs with flavour-changing couplings,
or a flavour-changing scalar operator
involving heavy fermions.

We first consider the QED$\times$QCD invariant EFT
below $m_W$, in the notation of \cite{C+C}, where
operators are added to the Lagrangian as
${\cal L}_{SM} \to {\cal L}_{SM} + 2\sqrt{2} G_F \Sigma C^\zeta_{Lor}
{\cal O}^\zeta_{\rm Lor}$, with $\zeta$ being  flavour indices, and
the subscript giving the Lorentz structure. 
As illustrated  in figure \ref{fig:matchconf},
 the scalar operators
${\cal O}^{\psi\psi}_{S,XX}, {\cal O}^{\psi\psi}_{S,XY}$ (see eqn  (\ref{opscaltau})),
match at $m_\psi$  onto ${\cal O}_{FF,X}$ and  ${\cal O}_{GG,X}$:
\bea
 \frac{C_{FF,X}}{v}
 & =&  -\frac{\a Q_\psi^2 N_{c,\psi} }{12\pi m_\psi(m_\psi)}
 (C^{\psi\psi}_{S,XX} + C^{\psi\psi}_{S,XY})
 \\
 \frac{C_{GG,X}}{v}
 & =& -\frac{\a_s(m_Q)}{24\pi m_Q (m_Q)}  (C^{QQ}_{S,XX} + C^{QQ}_{S,XY})
\label{FFGG}
\eea
where $Q\in \{c,b,t\}$.
We  focus on  $\psi \in \{\tau,c,b,t\}$  a heavy fermion,
because  the operators with $\psi \in \{e,u,d,s\}$ contribute
at tree level to
$\mec$ or $\meee$, and for $\psi = \mu$,
the operator contributes  at one loop to
$\meg$. It is interesting to pursue the
loop effects of these heavy-fermion scalars,  because
 the two heavy  fermions  make the operators difficult to probe 
directly in experiment.

The  ${\cal O}^{\psi\psi}_{S,XX} $ scalar operators (with 
the same chiral projector in both bilinears) contribute to the
dipole operator via ``Barr-Zee'' diagrams.
 The log$^2$-enhanced
 part is given by the one-loop RGEs of
QED \cite{megmW,PSI} as
\bea
\Delta C_{D,X} \approx  8 \frac{\alpha^2_e}{e(4\pi)^2} \left(
 C^{\tau \tau}_{S,XX} \frac{m_\tau}{m_\mu}
\ln^2\frac{m_W}{m_\tau} 
 +  \frac{4m_c}{3m_\mu}C^{cc}_{S,XX}
 \ln^2\frac{m_W}{m_c} 
+  \frac{m_b}{3m_\mu}C^{bb}_{S,XX}\ln^2\frac{m_W}{m_b} 
 \right)
\label{DCD}
\eea
where $C_{D,X}$ is the dipole coefficient at the experimental scale, and
the coefficients on the right are  evaluated at $m_W$.
Numerically, this is 
$$
\Delta C_{D,X} \approx 9 \times 10^{-6}  \left(
 245 C^{\tau \tau}_{S,XX}
  +  277C^{cc}_{S,XX}
+  117C^{bb}_{S,XX}  
 \right)
$$
where the current  MEG bound $BR(\meg) \leq 4.2 \times 10^{-13}$
gives $ C_{D,X} \lsim  10^{-8}$.
Comparing to eqns (\ref{FFGG}) and (\ref{bdnum}),  one sees that
the  scalar quark 
coefficients $C^{QQ}_{S,XX}$ give contributions to $\mec$
via the dipole and $GG$
operators that  are of the same order of magnitude and sign.
So the SINDRUMII  $\mec$ bound has  better sensitivity to these
operators \cite{C+C} than  the current  MEG bound. 
On the other hand, $C^{\tau \tau}_{S,XX}$ contributes principally to
$\mec$ via the dipole,  rather than the $FF$ operators,
so high precision would be required to see the $FF$
contribution,  and MEG has better sensitivity.

The  ${\cal O}_{S,XY}^{\psi \psi}$ operators can be Fierz-transformed to
vector operators   $-\frac{1}{2}(\overline{e} \gamma^\a \psi) 
(\overline{\psi} \gamma_\a \mu)$,  which contribute to $\meg$ at
two-loop in EFT\cite{PSI,Ciuchini:1993fk}, that is ${\cal O}(\a_e^2\ln)$:
\bea
\Delta C_{D,X} &\approx& \frac{\alpha^2_e}{e(4\pi)^2}\left[
 \frac{58}{9}C^{\tau\tau}_{V,YY}
 +  \frac{116}{9}\sum_{l=e,\m}C^{ll}_{V,YY}
+ \frac{64}{9}(C^{uu}_{V,YY} + C^{cc}_{V,YY})  +
 \frac{22}{9}\sum_{q=d,s,b}C^{qq}_{V,YY}
\right. \label{C2loop}\\
&&\left. 
-\frac{80}{9}(C^{uu}_{V,YX} + C^{cc}_{V,YX})
- \frac{14}{9}\sum_{q=d,s,b}C^{qq}_{V,YX}
- \frac{50}{9}\sum_{l=e,\m,\tau}C^{ll}_{V,YX}
+ 4\sum _{f=b,c,s,\tau} C_{S, YX}^{ff} \frac{Q_f^2 N_f m_f}{m_\mu}
\right]\ln\frac{m_W}{m_?} 
 \nonumber
\eea
where  the logarithm should be inside the bracket,
with a lower cutoff $\sim m_b\to m_\mu$
which depends on the operator.
For the  heavy quark  coefficients  $C_{S,XY}^{QQ}$,
  the  contribution
 to $\mec$ via  the ${\cal O}_{GG,X}$ operator
 is clearly larger than via the dipole or   ${\cal O}_{FF,X}$
 (see eqns (\ref{FFGG}),(\ref{bdnum})),
   giving the SINDRUMII search  the best  sensitivity.  

For the tau scalar  coefficient, eqn (\ref{C2loop})
 corresponds to
 $
\Delta C_{D,X}\simeq 2.9 \times 10^{-4} C_{S, YX}^{\tau\tau} 
$
which gives $\meg$ the current  best sensitivity
to this  coefficient \cite{C+C}.
For $\mec$, this contribution to the dipole can  be compared with
$
\Delta C_{FF,X}\simeq 0.019 C_{S, YX}^{\tau\tau}
$
from eqn (\ref{FFGG}).   Eqn (\ref{bdnum})
 then implies that the contribution of  $C_{S, YX}^{\tau\tau}$ to $(\mu{\rm Au}\to
e{\rm Au})$ via the dipole is an order of magnitude
larger than via the $FF$ operator, and
a similar dominance of the dipole contribution
arises in Aluminium. 
This can be understood diagrammatically,
where both   the contributions of 
$ {\cal O}_{S, YX}^{\tau\tau}$ 
to the dipole, and to the scalar
proton current,  arise at 2-loop
with a single log enhancement.
 However,
the contributions to the dipole benefit from a
$m_\tau/m_\mu$ enhancement.

\subsection{Of the sensitivity of $\mec$ to  flavour-changing
Higgs  interactions}

The discussion so far has been in the context of
QED$\times$QCD invariant operators below the weak scale.
However, since we assume $\LNP$ is large, it is relevant
to  translate to the SMEFT, where 
SU(2) invariance restricts   the operator basis 
to three scalar four-fermion operators at dimension six:
the $XX$ scalar for $u$-type quarks, and 
the $XY$ scalars for $d$-type quarks and
charged leptons.
There is  also a flavour-changing Higgs coupling,
which matches onto  ${\cal O}_{FF,X}$ and ${\cal O}_{GG,X}$
at the weak scale.
Including also the dipoles, these
operators appear in the SMEFT Lagrangian as 
\bea
\delta {\cal L}_{SMEFT} &= & \frac{1}{v^2}{\Big (}
C^{\mu e}_{EH}H^\dagger H \bar{\ell}_\mu H e
+
C^{e\mu}_{EW } y_\beta (\overline{\ell}_e  \tau^a H \sigma^{\mu \nu} e_\mu ) W^a_{\mu \nu}
+
C^{e\mu}_{EB }   y_\beta(\overline{\ell}_e H \sigma^{\mu \nu}  e_\mu ) B_{\mu \nu}  \label{SMEFT}\\
&&
+
 C^{e \tau\tau\mu}_{LE} (\overline{\ell}_e \gamma^\mu \ell_\tau ) 
(\overline{e}_\tau \gamma_\mu e_\mu)
+C^{\tau \mu e\tau}_{LE} (\overline{\ell}_\tau \gamma^\mu \ell_\mu ) 
(\overline{e}_e \gamma_\mu e_\tau) 
\nonumber \\
&&
+
C^{e\mu nn}_{ LEQU}  (\overline{\ell}_e^A  e_\mu ) \varepsilon_{AB}
(\overline{q}^B_n u_n )
+
C^{e\mu nn}_{LEDQ} (\overline{\ell}_e  e_\mu ) 
(\overline{d}_n q_n ) {\Big )}
+ h.c.\nonumber~~,
\eea
where the capitalized SU(2)  indices
are explicit when not contracted in the parentheses,
$\ell$ and $q$ are doublets, $u,d,e$ are singlets, 
flavour indices are superscripts, $n\in \{c,t,b\}$, and the operator
labels are according to \cite{polonais}.
The ${\cal O}^{e\mu}_{EW }$ and   ${\cal O}^{e\mu}_{EB }$
will combine to the dipole, the ${\cal O}_{LE }$
operators Fiertz to $XY$ scalar operators with a $\tau$
bilinear, and in the quark sector, ${\cal O}_{ LEQU}$
is a $YY$- scalar operator (same chiral projector twice),
whereas ${\cal O}_{ LEDQ}$ is $XY$.

Loop effects  between $\LNP$ and the weak scale can be
partially included via the RGEs of the
SMEFT.  Gauge boson loops can renormalize the coefficients,
and mix  the $C^{e\mu nn}_{ LEQU} $ coefficients into
the $u-$type tensor operator, and then to the dipole
(as occurs below $m_W$ for $YY$ scalars). 
Higgs exchange can mix these scalars into
vector four-fermion operators (to which there
could be better experimental sensitivity), but
for ${\cal O}_{ LEDQ}$ and  ${\cal O}_{LE }$,
this is negligible because suppressed by $\sim y_\mu y_\psi/(16\pi^2)$
($\psi \in\{\tau,b\}$).
We therefore  suppose that the coefficients in eqn
(\ref{SMEFT}) are given at the weak scale  $m_W$,
since the one-loop  RGEs above $m_W$ do not
appear to significantly mix the $XY$-scalars into more
experimentally accessible operators.

The coefficients  from eqn (\ref{SMEFT})
can  be matched
at $m_W$  onto those  of 
QED$\times$QCD-invariant scalar four-fermion operators,
relevant at low energy.
All the scalar operators below $m_W$ are generated
at tree level,  just that some 
arise due to Higgs exchange
with a flavour-changing  coupling from 
 the  ${\cal O}_{HE}$ operator, leading to correlations
in the coefficients.
One obtains \cite{megmW}
\bea
C_{D,R} (m_W)
& =&  c_W C^{e\mu}_{EB} (m_W) -s_W  C^{e\mu}_{EW} (m_W) +   C^{e\mu}_{EH}(m_W)  \left[ 
\frac{e\alpha y_t^2 }{8\pi^3y_\mu } 
\right] \label{Dmw}
\\
C^{ \tau \tau}_{S,RR}& =&  - \frac{m_\tau C^{e \mu  }_{EH} v}{m_h^2}\label{24} \\
C^{\tau \tau}_{S,LR}& =& -2 C_{LE}^{\tau \mu e \tau} - \frac{m_\tau C^{ \mu e *}_{EH} v}{m_h^2}\\
C^{ \tau \tau}_{S,RL}& =& -2 C_{LE}^{e \tau \tau \mu}
 - \frac{m_\tau C^{e \mu }_{EH} v}{m_h^2} \\
C^{ \tau \tau}_{S,LL}& =&  - \frac{m_\tau C^{ \mu e * }_{EH} v}{m_h^2}
\label{27}\\
C^{cc}_{S,LL}& =& C_{LEQU}^{* \mu e cc} -  \frac{m_{c} v}{m_h^2}  C^{ \mu  e *}_{EH} \label{39} \\
C^{ b b}_{S,LL}& =& -  \frac{m_{b} v}{m_h^2}  C^{ \mu  e *}_{EH} \label{40}\\
C^{ c c}_{S,RR}& =& C_{LEQU}^{e \mu  cc} -  \frac{m_{c} v}{m_h^2}  C^{ e \mu   }_{EH}\\
C^{ b b}_{S,RR}& =& -  \frac{m_{b} v}{m_h^2}  C^{e  \mu   }_{EH}\label{42}\\
C^{ c c}_{S,LR}& =&   -  \frac{m_{c} v}{m_h^2}  C^{  \mu e *  }_{EH}\\
C^{ b b}_{S,LR}& =& C_{LEDQ}^{* \mu e bb}  -  \frac{m_{b} v}{m_h^2}  C^{  \mu e *  }_{EH}\\
C^{ c c}_{S,RL}& =&  -  \frac{m_{c} v}{m_h^2}  C^{ e \mu   }_{EH}\\
C^{ b b}_{S,RL}& =& C_{LEDQ}^{e \mu  bb} -  \frac{m_{b} v}{m_h^2}  C^{ e \mu   }_{EH}
\label{46}\\
 \frac{C_{FF,R}}{v}
 & =&  -\frac{\a }{9\pi m_t} (C_{LEQU}^{e \mu  tt} -  \frac{2m_{t} v}{m_h^2}  C^{e \mu  }_{EH} ) \\
 \frac{C_{GG,R}}{v}
 & =& -\frac{\a_s}{24\pi m_t} (C_{LEQU}^{e \mu  tt} -  \frac{2m_{t} v}{m_h^2}  C^{e \mu   }_{EH} )
 \eea
where $s_W = \sin \theta_W$,  all the masses and couplings  are  running,
and are   evaluated at the
weak scale. The  two-loop Barr-Zee  diagrams  involving top and
$W$ loops were included in the matching to the dipole,
and the top loop matching the scalar operators onto $FF$
and $GG$ was included for these operators.

These coefficients then run down
to the experimental scale with the
RGEs
of QED$\times$QCD (see, eg,
\cite{PSI}).
QCD effects are
numerically significant, although
they only renormalize the coefficients\footnote{QCD
can also mix  ${\cal O}_{GG,X}$ to
${\cal O}_{S,XL}+  {\cal O}_{S,XR}$ by attaching the gluons
to heavy quark line with a mass insertion. But we
do not include this, because the scalar operators
always have to be matched back to ${\cal O}_{GG,X}$
in order to contribute to $\mec$.}.
The scalar quark  operators
run like  quark masses, and the operator
${\cal O}_{GG,X}$ runs like the gluon kinetic term,
{ which is accounted for by
the wave function renormalization of the gluons.
So the running parameters
in the coefficient  are evaluated at the matching scale.  }

Retaining only the contribution of the flavour-changing Higgs
couplings, we obtain
\bea
C^{e\mu}_{D,R} &\simeq &  \left[C^{e\mu}_{e\g}  + \frac{e \alpha_e y_t^2}{8 \pi^3y_\mu}
C^{e\mu}_{EH} \right] \left[ 1 -\frac{4 \a_e}{\pi} \ln \left (\frac{m_W}{m_\mu}
\right)\right] +...
\label{Cinput}\\
C_{GG,R} &=&  \frac{v^2}{12\pi m_h^2}
C^{e\mu}_{EH}  \left[ \alpha_s (m_t) + \alpha_s (m_b) + \alpha_s (m_c)\right]
+...
\nonumber\\
C_{FF,R} &=&  \frac{\alpha_e  v^2}{6 \pi m_h^2}
C^{e\mu}_{EH}  \left[  \frac{9}{3} + 1\right] +...
 \nonumber
\eea
where on the right appear  SMEFT coefficients evaluated
at $m_W$, and  the coefficients on the left  can be input into
the  rate for $\mec$. 
Here $C^{e\mu}_{e\g} = c_W C^{e\mu}_{EB}  -s_W  C^{e\mu}_{EW}$.

Combining with eqn (\ref{bdnum}), one sees that
the contribution of the  LFV Higgs  interactions
 ${\cal O}^{e  \mu}_{EH}$
via  ${\cal O}_{GG,X}$  is of
 opposite sign and $\frac{1}{3}$ the magnitude of the
 dipole contribution. The contribution
 via the light quark ($u,d,s$) scalar operators is
 slightly smaller than the $GG$ contributions and of same sign,
 which worsens the sensitivity
 of $\mec$ to  ${\cal O}_{EH}$
\footnote{This cancellation
 is more effective
 for light  targets like Aluminium or Titanium.}.
 Including both effects, $\mu{\rm Au} \to e+{\rm Au}$
cannot see
\beq
C^{e  \mu}_{EH} \leq  4.7\times 10^{-5} 
\eeq
whereas including only the dipole would give a sensitivity
of $\lsim 1.6\times 10^{-5}$.

\section{Numerical values of overlap integral}
\label{app:FA}

We show the numerical values of $F_A$, defined in Sec.~\ref{sec:caln}. 
A few nuclei are modeled by both the 2pF and 3pG distributions 
\eqref{Eq:rho} \cite{Jager1974}, in which case we give the results with 
the latest distribution: 3pG for $Z=16, 28, 38, 40, 42, 50, 56$, and $83$, 
and 2pF for other nuclei.

\begin{table}[ht]
\centering
 \begin{tabular}{lclc}
 \hline
 Nucleus & $F_A\times 10^4$ & Nucleus & $F_A\times 10^4$ \\
\hline
$_{9}^{19}$F & 1.5 & $_{40}^{90}$Zr & $-0.67$ \vspace*{1mm} \\
$_{10}^{20}$Ne & 1.7 & $_{41}^{93}$Nb & $-1.3$ \vspace*{1mm} \\
$_{12}^{24}$Mg & 2.7 & $_{42}^{98}$Mo & $-4.0$ \vspace*{1mm} \\
$_{13}^{27}$Al & 3.8 & $_{48}^{114}$Cd & $-13$ \vspace*{1mm} \\
$_{14}^{28}$Si & 4.5 & $_{49}^{115}$In & $-11$ \vspace*{1mm} \\
$_{15}^{31}$P & 4.3 & $_{50}^{120}$Sn & $-15$ \vspace*{1mm} \\
$_{16}^{32}$S & 5.0 & $_{51}^{121}$Te & $-14$ \vspace*{1mm} \\
$_{18}^{40}$Ar & 5.6 & $_{56}^{138}$Ba & $-25$ \vspace*{1mm} \\
$_{20}^{40}$Ca & 7.4 & $_{57}^{139}$La & $-25$ \vspace*{1mm} \\
$_{22}^{48}$Ti & 7.1 & $_{60}^{142}$Nd & $-29$ \vspace*{1mm} \\
$_{23}^{51}$V & 7.1 & $_{62}^{152}$Sm & $-32$ \vspace*{1mm} \\
$_{24}^{52}$Cr & 7.2 & $_{67}^{165}$Ho & $-40$ \vspace*{1mm} \\
$_{25}^{55}$Mn & 8.3 & $_{73}^{181}$Ta & $-46$ \vspace*{1mm} \\
$_{26}^{56}$Fe & 7.5 & $_{74}^{184}$W & $-51$ \vspace*{1mm} \\
$_{27}^{59}$Co & 7.5 & $_{79}^{197}$Au & $-61$ \vspace*{1mm} \\
$_{28}^{58}$Ni & 7.8 & $_{82}^{208}$Pb & $-63$ \vspace*{1mm} \\
$_{29}^{63}$Cu & 6.6 & $_{83}^{209}$Bi & $-65$ \vspace*{1mm} \\
$_{30}^{64}$Zn & 6.2 & $_{90}^{232}$Th & $-74$ \vspace*{1mm} \\
$_{38}^{88}$Sr & $-0.26$ & $_{92}^{238}$U & $-75$ \vspace*{1mm} \\
$_{39}^{89}$Y & $-0.37$ & - & - \\
\hline
\end{tabular}
 \caption{$F_A$ for each nucleus.
   \label{tab:fa} }
\end{table}

\end{document}